\documentclass{article}

\usepackage[preprint]{neurips_2019}
\usepackage{amsmath}
\numberwithin{equation}{section}
\usepackage{CJKutf8}
\usepackage[T1]{fontenc}    
\usepackage{hyperref}       
\usepackage{url}            
\usepackage{booktabs}       
\usepackage{amsfonts}       
\usepackage{nicefrac}       
\usepackage{microtype}      
\usepackage{graphicx}
\usepackage{epstopdf}
\usepackage{setspace}
\usepackage{rotating}
\usepackage{subfig}
\usepackage{verbatim}
\usepackage{amssymb}
\usepackage{amsmath}
\usepackage{algorithm2e}
\usepackage{verbatim} 
\usepackage{epstopdf}
\usepackage{setspace}
\usepackage{amsthm}
\usepackage{rotating}
\usepackage[T1]{fontenc}
\usepackage{subfig}

\usepackage{comment}
\usepackage{hyperref}
\usepackage{tikz,xcolor,hyperref}
\usepackage{caption} 
\usepackage{hyperref}
\usepackage{setspace}
\usepackage{soul}
\usepackage{lineno}
\usepackage{listings}
\usepackage{color}
\usepackage{verbatim}
\definecolor{mygreen}{rgb}{0,0.6,0}
\definecolor{mygray}{rgb}{0.5,0.5,0.5}
\definecolor{mymauve}{rgb}{0.58,0,0.82}
\definecolor{lime}{HTML}{A6CE39}
\DeclareRobustCommand{\orcidicon}{
	\begin{tikzpicture}
	\draw[lime, fill=lime] (0,0) 
	circle [radius=0.16] 
	node[white] {{\fontfamily{qag}\selectfont \tiny ID}};
	\draw[white, fill=white] (-0.0625,0.095) 
	circle [radius=0.007];
	\end{tikzpicture}
	\hspace{-2mm}
}
\foreach \x in {A, ..., Z}{%
	\expandafter\xdef\csname orcid\x\endcsname{\noexpand\href{https://orcid.org/\csname orcidauthor\x\endcsname}{\noexpand\orcidicon}}
}

\usepackage[compact]{titlesec} 
\title{Inverted repeats in coronavirus SARS-CoV-2 genome and implications in evolution}
\author{%
	Changchuan Yin \orcidA \thanks{Correspondence author, cyin1@uic.edu}\\
	Department of Mathematics, Statistics, and Computer Science \\
	University of Illinois at Chicago \\
	Chicago, IL 60607 \\
	USA\\
	\AND
	Stephen S.-T. Yau \orcidB \thanks{Correspondence author, yau@uic.edu}\\
	Department of Mathematical Sciences\\
	Tsinghua University\\
	Beijing 100084 \\
	China\\
}
\begin{document}
\maketitle
\begin{abstract}
The coronavirus disease (COVID-19) pandemic, caused by the coronavirus SARS-CoV-2, has caused 60 millions of infections and 1.38 millions of fatalities. Genomic analysis of SARS-CoV-2 can provide insights on drug design and vaccine development for controlling the pandemic. Inverted repeats in a genome greatly impact the stability of the genome structure and regulate gene expression. Inverted repeats involve cellular evolution and genetic diversity, genome arrangements, and diseases. Here, we investigate the inverted repeats in the coronavirus SARS-CoV-2 genome. We found that SARS-CoV-2 genome has an abundance of inverted repeats. The inverted repeats are mainly located in the gene of the Spike protein. This result suggests the Spike protein gene undergoes recombination events, therefore, is essential for fast evolution. Comparison of the inverted repeat signatures in human and bat coronaviruses suggest that SARS-CoV-2 is mostly related SARS-related coronavirus, SARSr-CoV/RaTG13. The study also reveals that the recent SARS-related coronavirus, SARSr-CoV/RmYN02, has a high amount of inverted repeats in the spike protein gene. Besides, this study demonstrates that the inverted repeat distribution in a genome can be considered as the genomic signature. This study highlights the significance of inverted repeats in the evolution of SARS-CoV-2 and presents the inverted repeats as the genomic signature in genome analysis.

\end{abstract}
\textbf{{\large keywords}}: COVID-19, SARS-CoV-2, 2019-nCoV, coronavirus, inverted repeat, genome, evolution

\section{Introduction}
\label{Introduction}
\subsection{Human coronaviruses}
The novel human coronavirus SARS-CoV-2 (formerly, 2019-nCoV) first emerged in Wuhan, China, in December 2019, the causative agent for Coronavirus Disease-2019 (COVID-19) pandemic, has claimed 1.38 million mortality in the globe as of Nov.24, 2020 \citep{owidcoronavirus}. Understanding the molecular structure and evolution of SARS-CoV-2 genome is of urgency for tracing the origin of the virus and provides insights on vaccine development and drug design for controlling the current COVID-19 pandemic. 

Human coronaviruses (CoVs) are common viral respiratory pathogens that cause mild to moderate upper-respiratory tract illnesses. Two common CoVs, 229E, and OC43 were identified in 1965 and can cause the common cold. Four typical human CoVs found in recent years are Severe Acute Respiratory Syndrome Coronavirus (SARS-CoV) in 2002, NL63 in 2004, HKU1 in 2005, and Middle East respiratory syndrome coronavirus (MERS-CoV) in 2012. Among these human CoVs, SARS-CoV and MERS-CoV are highly pathogenic and caused severe and fatal infections. MERS symptoms are very severe, usually including fever, cough, and shortness of breath which often progress to pneumonia. About 30\% with MERS had died. SARS symptoms often include fever, chills, and body aches which usually progressed to pneumonia. About 10\% with SARS-CoV had died. The current coronavirus SARS-CoV-2, which causes a worldwide COVID-19 pandemic, is milder than SARS-CoV, but can cause severe syndromes and fatality in people with cardiopulmonary disease, people with weakened immune systems, infants, and older adults.

SARS-CoV-2 is a beta coronavirus, like MERS-CoV and SARS-CoV. All three of these coronaviruses have their origins in bats. Yet the zoonotic origin of SARS-CoV-2 is still unconfirmed. \cite{zhou2020pneumonia,zhou2020pneumoniaB} 's study showed that the bat SARS-related coronavirus strain SARSr-CoV/RaTG13, identified from a bat \textit{Rhinolophus affinis} in Yunnan province, China, in July 2012, shares 96.2\% nucleotide identity. A recent study identified a new SARSr-CoV/RmYN02 (2019) from \textit{Rhinolophus malayanus}, which is closely related to SARS-CoV-2 \citep{zhou2020novel}. SARSr-CoV/RmYN02 shares 93.3\% nucleotide identity with SARS-CoV-2 and comprises natural insertions at the S1/S2 cleavage site of the Spike protein. The unique S1/S2 cleavage in the Spike protein in SARS-CoV-2 may confer the zoonotic spread of SARS-CoV-2. However, the originating relationship among these CoVs is not entirely clear.

\subsection{Coding structures of SARS-CoV-2 genome}
SARS-CoV-2 coronavirus contains a linear single-stranded positive RNA genome (Fig.1). The SARS-CoV-2 RNA genome of 29.9kb has a total of 11 genes with 11 open reading frames (ORFs) \citep{yoshimoto2020proteins}, consisting of the leader sequence (5'UTR), the coding regions, and 3'UTR pseudoknot stem-loop \citep{wu2020new}. The coding regions include ORF1ab and genes encoding 16 non-structural proteins \citep{finkel2020coding} and structural proteins (spike (S), envelope (E), membrane (M), and nucleocapsid (N)) \citep{gordon2020sars}, and several accessory proteins. 

ORF1ab encodes replicase polyproteins required for viral RNA replication and transcription \citep{chen2020emerging,cavasotto2020functional}. Nonstructural protein 1 (nsp1) likely inhibits host translation by interacting with 40S ribosomal subunit, leading to host mRNA degradation through cleavage near their 5’UTRs. Nsp 1 promotes viral gene expression and immunoevasion in part by interfering with interferon-mediated signaling. Nonstructural protein 2 (nsp2) interacts with host factors prohibitin 1 and prohibitin 2, which are involved in many cellular processes including mitochondrial biogenesis. The third non-structural protein (nsp3) is Papain-like proteinase. Nsp3 is an essential and the largest component of the replication and transcription complex. The Papain-like proteinase cleaves non-structural proteins 1-3 and blocks the host's innate immune response, promoting cytokine expression \citep{serrano2009nuclear,lei2018nsp3}. Nsp4 encoded in ORF1ab is responsible for forming double-membrane vesicle (DMV). The other non-structural proteins are 3CLPro protease (3-chymotrypsin-like proteinase, 3CLpro) and nsp6. 3CLPro protease is essential for RNA replication. The 3CLPro proteinase accounts for processing the C-terminus of nsp4 through nsp16 in coronaviruses \citep{anand2003coronavirus}. Together, nsp3, nsp4, and nsp6 can induce DMV \citep{angelini2013severe}. 

SARS-coronavirus has a unique RNA replication facility, including two RNA-dependent RNA polymerases (RNA pol). The first RNA polymerase is a primer-dependent non-structural protein 12 (nsp12), and the second RNA polymerase is nsp8, nsp8 has the primase capacity for \textit{de novo} replication initiation without primers \citep{te2012sars}. Nsp7 and nsp8 are essential proteins in the replication and transcription of SARS-CoV-2. Nsp7 is responsible for nuclear transport. The SARS-coronavirus nsp7-nsp8 complex is a multimeric RNA polymerase for both \textit{de novo} initiation and primer extension \citep{prentice2004identification, te2012sars}. Nsp8 also interacts with ORF6 accessory protein. The nsp9 replicase protein of SARS-coronavirus binds RNA and interacts with nsp8 for its functions \citep{sutton2004nsp9}. Helicase (nsp13) possesses helicase activity, thus catalyzing the unwinding dsRNA or structured RNA into single strands. Importantly, nsp14 may function as a proofreading exoribonuclease for virus replication, hence, SARS-CoV-2 mutation rate remains low.

\begin{figure}[tbp]
	\centering
	{\includegraphics[width=4.0in]{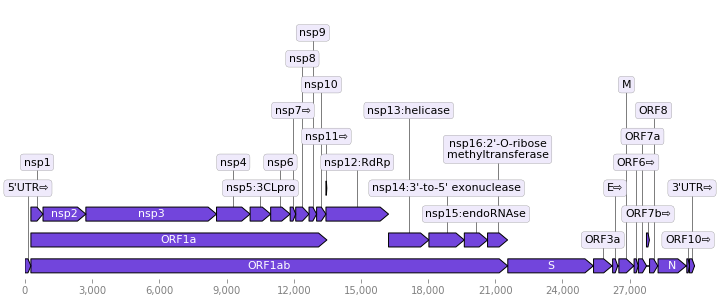}}\quad%
	\caption{The structural diagram of SARS-CoV-2 genome (GenBank: NC\_045512). The diagram of SARS-CoV-2 genome was made using DNA Feature Viewer \citep{zulkower2020dna}.}
	\label{fig:sub1}
\end{figure}

Furthermore, the SARS-CoV-2 genome encodes several structural proteins. The structural proteins possess much higher immunogenicity for T cell responses than the non-structural proteins \citep{li2008t}. The structural proteins include spike (S), envelope (E), membrane protein (M), and nucleoprotein (N) \citep{marra2003genome,ruan2003comparative}. The Spike glycoprotein has two domains S1 and S2. Spike protein S1 attaches the virion to the host cell membrane through the receptor ACE2, initiating the infection \citep{wan2020receptor, wong2004193}. After being internalized into the endosomes of the cells, the S glycoprotein is then cleaved by cathepsin CTSL. The spike protein domain S2 mediates fusion of the virion and cellular membranes by acting as a class I viral fusion protein. Especially, the spike glycoprotein of coronavirus SARS-CoV-2 contains a furin-like cleavage site \citep{coutard2020spike}. Recent study indicates that SARS-CoV-2 is more infectious than SARS-CoV according to the changes of S protein-ACE2 binding affinity \citep{chen2020mutations}. The envelope (E) protein interacts with membrane protein M in the budding compartment of the host cell. The M protein holds dominant cellular immunogenicity \citep{liu2010membrane}. Nucleoprotein (ORF9a) packages the positive-strand viral RNA genome into a helical ribonucleocapsid (RNP) during virion assembly through its interactions with the viral genome and a membrane protein M \citep{he2004characterization}. Nucleoprotein plays an important role in enhancing the efficiency of subgenomic viral RNA transcription and viral replication. 

\subsection{Non-coding structures of the SARS-CoV-2 genome}
In addition to the coding regions, SARS-CoV-2 genome contains hidden structures that can retain genome stability, regulate gene replication and expression, and control virus life cycles. The non-coding genome structures include leader sequences, transcriptional regulatory sequences (TRS), G-quadruplex structures, frame-shifting regions, and repeats. The first non-coding structure is the 5' leader sequence of about 265 bp is the unique characteristic in coronavirus replication and plays critical roles in the gene expression of coronavirus during its discontinuous sub-genomic replication \citep{li2005sirna}. 

SARS-CoV-2 contains G-quadruplex structures \citep{ji2020discovery}. It is well established that sequences with G-blocks (adjacent runs of Guanines) can potentially form non-canonical G-quadruplex (G4) structures \citep{choi2011conformational,metifiot2014g}. The G4 structures are formed by stacking two or more G-tetrads by Hoogsteen hydrogen bonds and often are the sites of genomic instability, serving one or more biological functions \citep{bochman2012dna}. 

An inverted repeat is a single-stranded sequence of nucleotides followed by downstream its reverse complement downstream. The intervening sequence between the initial sequence and the reverse complement is called a spacer. When the spacer sequence is zero, the inverted repeat is called a palindrome. For example, the inverted repeat, 5'-ATTCGCGAAT-3' is a palindrome, the palindrome-first sequence is 5'-ATTCG-3', and the palindrome-second sequence is 5'-CGAAT-3'. When the spacer in an inverted repeat is non-zero, the repeat is generally inverted. In a generally inverted repeat, we still denote the initial sequence as a palindrome-first sequence and the downstream reverse complement as a palindrome-second sequence. For example, in the general inverted repeat, 5'-TTTAGGT...ACCTAAA-3', the palindrome-first sequence is 5'-TTTAGGT-3', and the palindrome-second sequence is 5'-ACCTAAA-3'. Through self-complementary base pairing, an inverted repeat can form a stem-loop (hairpin) structure in an RNA molecule, where the palindrome-first and palindrome-second sequences make a stem, and the spacer sequence makes a loop. It should be noted that an inverted repeat may not have perfect complementary base pairing in palindrome-first and palindrome-second sequences, so the stem formed by an imperfect inverted repeat can have mismatches, insert, or deletions. Inverted repetitive sequences are principal components of the archaeal and bacterial CRISPR-CAS systems \citep{mojica2005intervening}, which function as adaptive antiviral defense systems.

Inverted repeats have important biological functions in viruses. Inverted repeats delimit the boundaries in transposons in genome evolution and form stem-loop structures in retaining genome instability and flexibility. Inverted repeats are described as hotspots of eukaryotic and prokaryotic genomic instability\citep{voineagu2008replication}, replication \citep{pearson1996inverted}, and gene silencing \citep{selker1999gene}. Therefore, inverted repeats involve cellular evolution and genetic diversity, mutations, and diseases.

Despite the paramount roles of the non-coding structures, the non-coding structures are not immediately visible as the coding regions. This study is to identify one of the crucial non-coding structures, inverted repeats in SARS-CoV-2 genome, and investigate the cohort of the inverted repeats and the virus evolution. 

\section{Materials and methods}
\subsection{Identification of inverted repeats}
The complete genomes of coronaviruses were scanned for inverted repeats using Palindrome analyzer \citep{brazda2016palindrome}. Palindrome analyzer (http://bioinformatics.ibp.cz/) is a web-based server for retrieving palindromic and inverted repeats in DNA or RNA sequences. Palindrome server describes the features of  inverted repeats including similarity analysis, localization, and visualization.

\subsection{Inverted repeat analysis}
To ensure consistency in comparing coronavirus genomes, we only extracted the inverted repeats with the perfect complementary base pairing of the palindrome-first and palindrome-second sequences. Noted that a short inverted repeat of length $P$ can be inside a long inverted repeat of length $Q$ ($Q>P$), in this case, we only extracted the inverted repeats of length $Q$ and excluded the inverted repeat of length $P$.

The retrieved inverted repeats were mapped on the protein genes in a genome according to the positions of the palindrome-first and palindrome-second sequences of the inverted repeats.

The distributions of inverted repeats on protein genes in the different genomes are assessed by the Wasserstein distance, known as the earth mover’s distance. The Wasserstein distance corresponds to the minimum amount of work required to transform one distribution into the other. The $p-th$ Wasserstein distance between two probability distributions $\mu$ and $\nu$ is defined as follows \citep{vallender1974calculation}, 

\[
W_p (\mu ,\nu ) = \left( {\mathop {\inf }\limits_{\pi  \in \Gamma (\mu ,\nu )} \int_{\mathbb{R} \times \mathbb{R}} {\left| {x - y} \right|d\pi (x,y)} } \right)^{1/p} 
\]

, where $\Gamma (\mu ,\nu )$ denotes the set of probability distributions on $\mathbb{R} \times \mathbb{R}$ with marginals $\mu$ and $\nu$. 

\subsection{Genome data}
The following complete genomes of SARS-CoVs and SARS-related coronaviruses (SARSr-CoVs) were downloaded  from NCBI GenBank: SARS-CoV-2 (GenBank: NC\_045512.2) \citep{wu2020new}, SARS-COV/BJ01 (GenBank: AY278488), SARSr-CoV/RaTG13 (GenBank: MN996532) \citep{zhou2020pneumonia}, SARSr-CoV/RmYN02 (GISAID: EPI\_ISL\_412977) \citep{zhou2020novel,shu2017gisaid}, and MERS-CoV (GenBank: NC\_019843) \citep{zaki2012isolation}. 

\section{Results}
\subsection{Inverted repeats in SARS-CoV-2 genome}
Long inverted repeats are deemed to greatly influence the stability of the genomes of various organisms. The longest inverted repeats identified in SARS-CoV-2 genome is 15 bp sequence, the palindrome-first sequence 5'-ACTTACCTTTTAAGT-3' is at 8474-8489 (nsp3 gene), and the palindrome-second sequence 5'-ACTTAAAAGGTAAGT-3' is at 13295-13310 (nsp10 gene). The repeats of 11-15 bp are predominantly located in the gene of the Spike (S) protein (Fig.2(a) and (b)). The other three protein genes (nsp3, RdRp, and N protein) are also enriched with long inverted repeats.

Long inverted repeats often contribute to the stability of a genome because of stable stems formed by the long inverted repeats. The results also suggest the recombinations took place at the gene of the Spike protein during evolution. Together, four protein genes (S, nsp3, RdRp, and N protein) of abundant inverted repeats are evolving dramatically and are critical for virus survival, therefore, can be the pharmaceutical targets \citep{gao2020machine}.

\begin{figure}[tbp]
	\centering
	\subfloat[]{\includegraphics[width=3.75in]{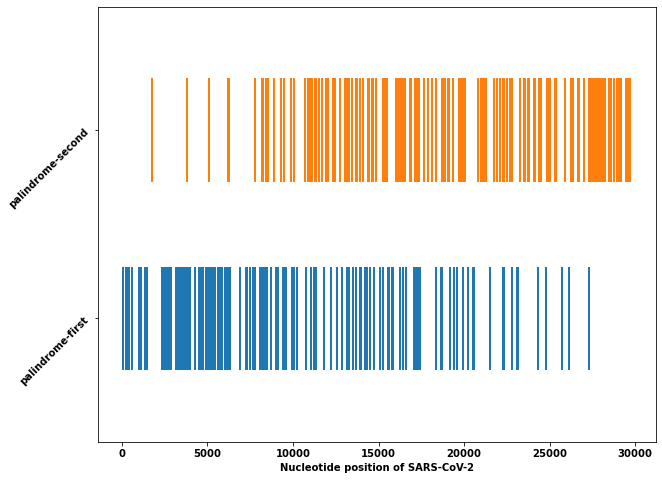}}\quad
	\subfloat[]{\includegraphics[width=3.75in]{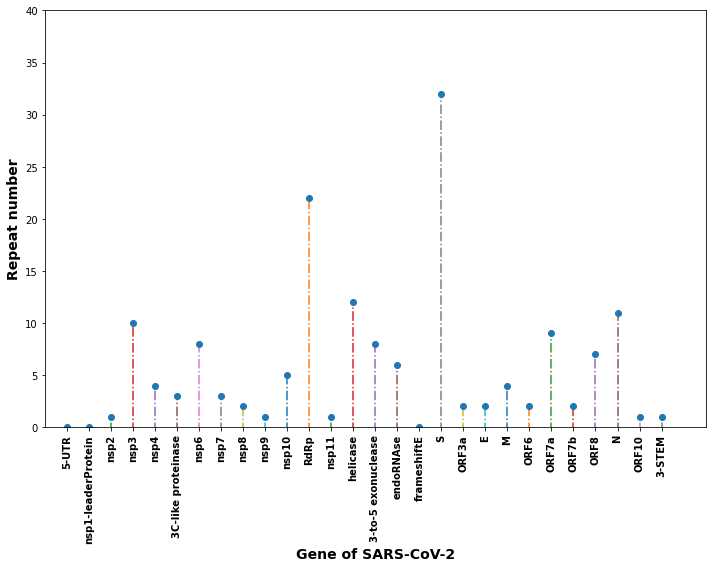}}\quad
	\caption{Distributions of inverted repeats consisting of first half sequences and second sequences on SARS-CoV-2 genome (NC\_045512). (a) Inverted repeats of 11-15 bp. (b) Repeat numbers of inverted repeats of 12-15 bp in the protein genes of the genome. In (b), the repeat numbers are counted by both palindrome-first and palindrome-second sequences.}
\end{figure}

The relation of virus genomes may provide insights on the zoonotic origin and evolution of the viruses. To examine the close relevance of human and bat CoVs, we evaluate and compare the distributions of inverted repeats of 11-15 bp in four CoV genomes: SARS-CoV-2 (Fig.2(a)), SARS-CoV (Fig.3(a)), MERS-CoV (Fig.4(a)) SARSr-CoV/RaTG13 (Fig.5(a)), and SARSr-CoV/RmYN02 (Fig.6(a)). The repeat numbers of the inverted repeats of 11-15 bp on each protein gene in the genomes are shown in Fig.2(b), Fig.3(b), Fig.4(b), Fig.5(b), and Fig.6(b). The repeat numbers are counted by both the palindrome-first and palindrome-second sequences of the inverted repeats. Taking account of the inverted repeats of wide ranges 8-15 bp, we computed the pairwise Wasserstein distances of the repeat numbers of protein genes in three closely related SARSr-CoVs: the distance between SARS-CoV-2 and SARSr-CoV/RaTG1 is 6.8571, the distance between SARS-CoV-2 and SARSr-CoV/RmYN02 is 5.7143, and the distance between SARSr-CoV/RaTG1 and SARSr-CoV/RmYN02 is 6.3571. Therefore, we conclude that SARS-CoV-2 strain is more closely related to SARSr-CoV/RaTG1 (2013) than SARSr-CoV/RmYN02 (2019). Both SARS-CoV-2 and SARSr-CoV/RmYN02 may evolve from SARSr-CoV/RaTG1. We also observe that the Spike protein gene in SARSr-CoV/RmYN02 (Fig.6(b)) have more long inverted repeats than the counterparts of SARS-CoV-2 (Fig.2(b)) and SARSr-CoV/RaTG1 (Fig.5(b)). Unsurprisingly, the Spike protein in  SARSr-CoV/RmYN02 contains natural insertions at the S1/S2 cleavage site. This cleavage site may originate from some recombination events of the Spike genes as the result of inverted repeats. 

\begin{figure}[tbp]
	\centering
	\subfloat[]{\includegraphics[width=3.75in]{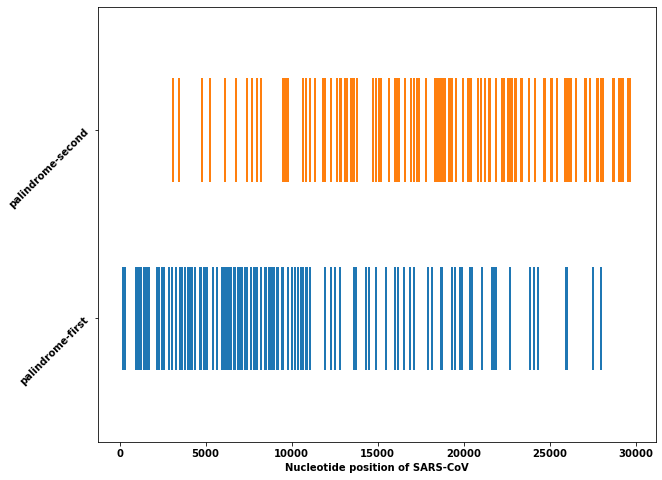}}\quad
	\subfloat[]{\includegraphics[width=3.75in]{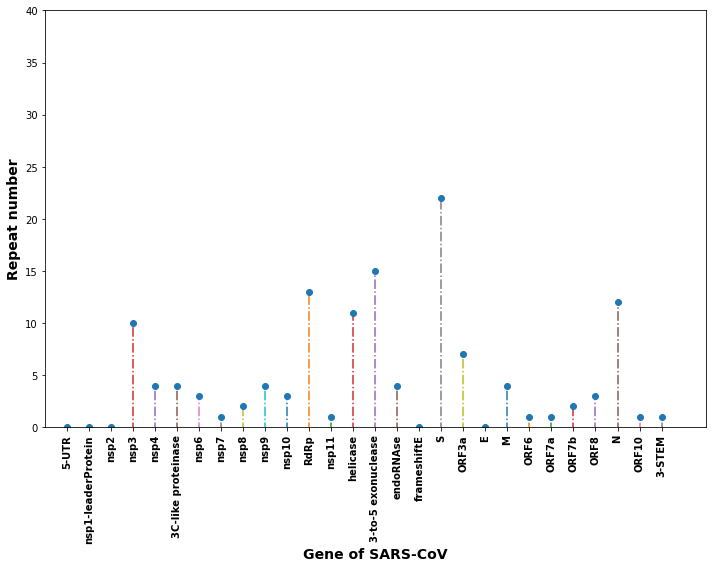}}\quad
	\caption{Distributions of inverted repeats consisting of palindrome-first and palindrome-second sequences on SARS-CoV genome (AY278488). (a) Inverted repeats of 11-15 bp. (b) Repeat numbers of inverted repeats of 12-15 bp in the protein genes of the genome.}
\end{figure}

\begin{figure}[tbp]
	\centering
	\subfloat[]{\includegraphics[width=3.75in]{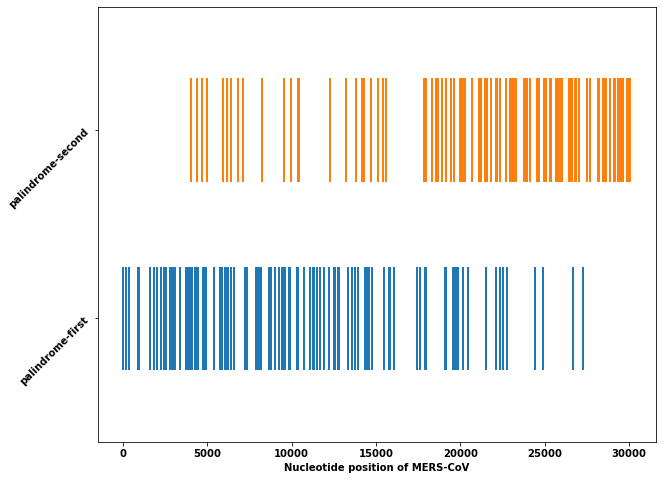}}\quad
	\subfloat[]{\includegraphics[width=3.75in]{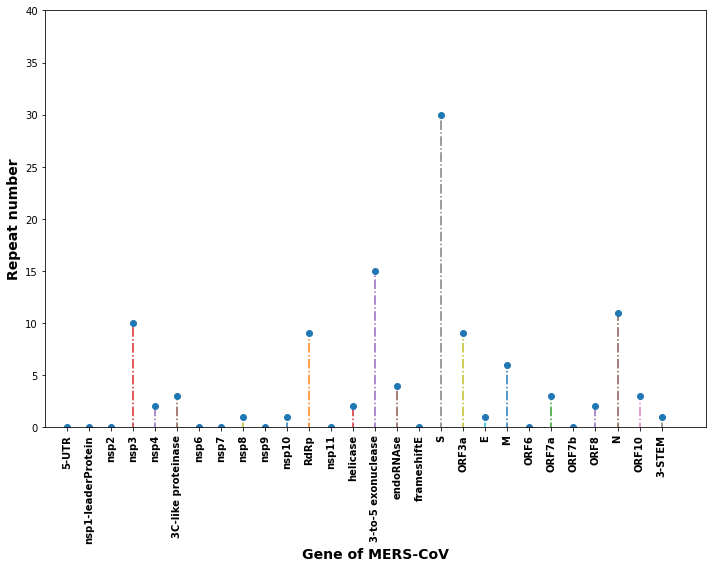}}\quad
	\caption{Distributions of inverted repeats consisting of palindrome-first and palindrome-second sequences on MERS-CoV genome (NC\_019843). (a) Inverted repeats of 11-15 bp. (b) Repeat numbers of inverted repeats of 12-15 bp in the protein genes of the genome.}
\end{figure}

\begin{figure}[tbp]
	\centering
	\subfloat[]{\includegraphics[width=3.75in]{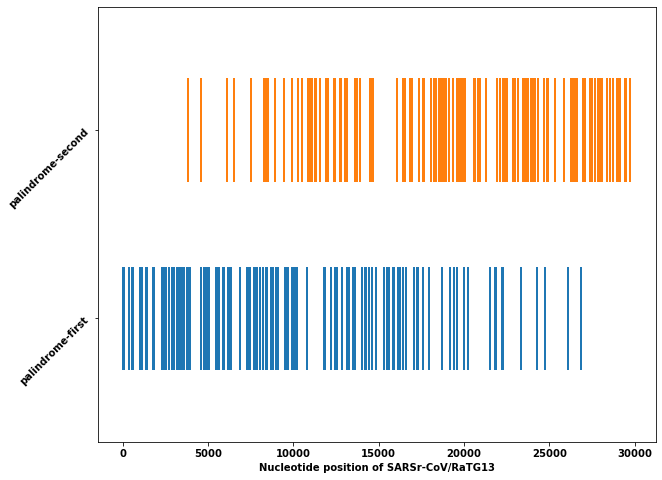}}\quad
	\subfloat[]{\includegraphics[width=3.75in]{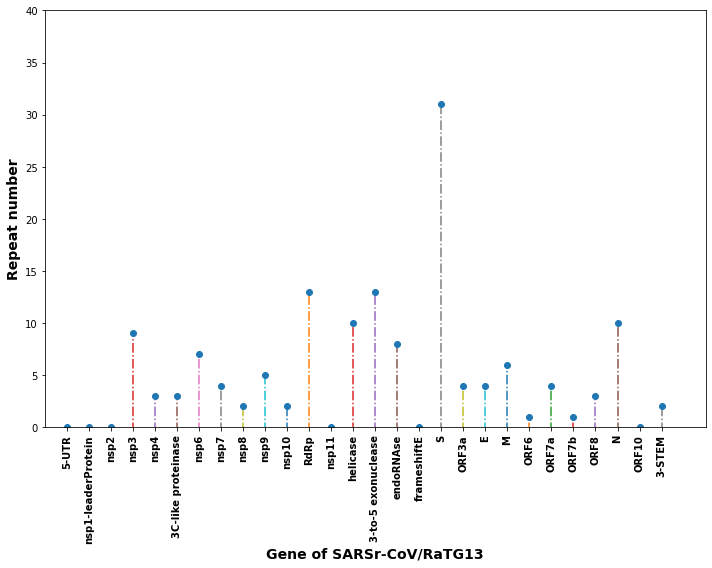}}\quad
	\caption{Distributions of inverted repeats consisting of palindrome-first and palindrome-second sequences on SARSr/RaTG13 genome (MN996532). (a) Inverted repeats of 11-15 bp. (b) Repeat numbers of inverted repeats of 12-15 bp in the protein genes of the genome.}
\end{figure}

\begin{figure}[tbp]
	\centering
	\subfloat[]{\includegraphics[width=3.75in]{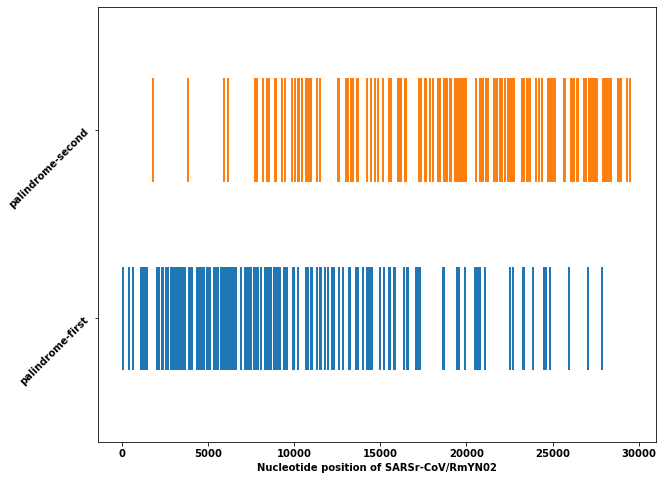}}\quad
	\subfloat[]{\includegraphics[width=3.75in]{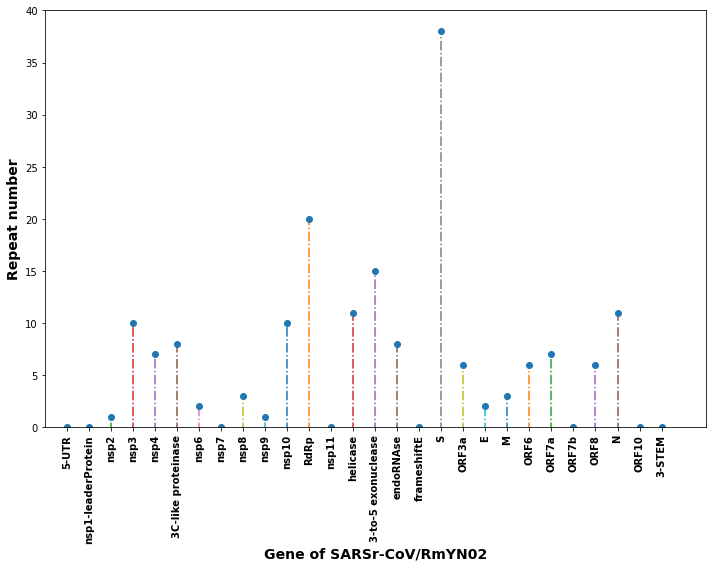}}\quad
	\caption{Distributions of inverted repeats consisting of palindrome-first and palindrome-second sequences on SARSr-CoV/RmYN02 genome (EPI\_ISL\_412977). (a) Inverted repeats of 11-15 bp. (b) Repeat numbers of inverted repeats of 12-15 bp in the protein genes of the genome.}
\end{figure}

The total frequencies of inverted repeats of different lengths in the human and bat CoVs also suggest that SARS-CoV-2 is closely related SARSr-CoV/RaTG13 (Fig.7). Notedly, Fig. 7 shows that the inverted repeats of all lengths are increasing from SARS-CoV (in 2003) to SARS-CoV-2 (in 2019).

From these repeat analyses, we may infer that during evolution, the recombinations may occur and produce accumulative inverted repeats under natural selection. We see that recombinations can be one of the driven forces for fast evolution.  

\begin{figure}[tbp]
	\centering
	{\includegraphics[width=4.0in]{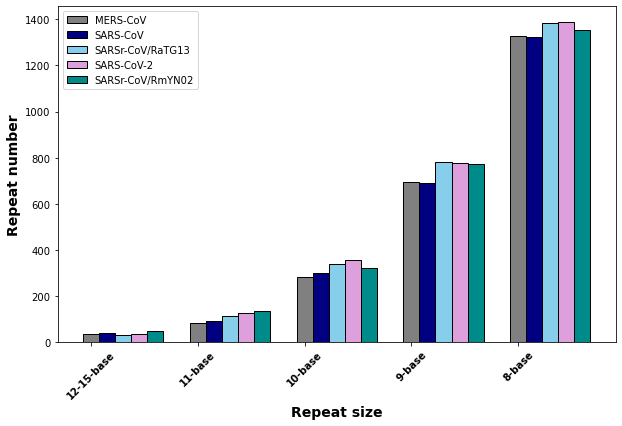}}\quad%
	\caption{Frequencies of inverted repeats of different lengths in the coronavirus genomes: SARS-CoV-2, SARS-CoV, MERS-CoV, SARSr-CoV/RaTG13, and SARSr-CoV/RmYN02. The repeat numbers are counted by palindrome-first sequences only.}
	\label{fig:sub1}
\end{figure}

\section{Discussions}
The COVID-19 pandemic has caused substantial health emergencies and economic stress in the world. Vaccine development is critical to mitigating the pandemic. The facts revealed in this study that three proteins nsp3, RdRp, and the Spike protein are rich with inverted repeats suggest that these three proteins are functional significance for virus survivals, and shall be the targets of drug design and vaccine development.

If we relax the matching pairs in the inverted repeats, we expect that much longer inverted repeats can be identified, and the number of inverted repeats in the virus genome will be increased significantly. The imperfect inverted repeats are the natural forms of the repeats to maintain the genome structures. Because the perfect inverted repeat distribution and types in a genome are unique and extracting the perfect inverted repeats are parameter-free, the perfect inverted repeats can be considered as the genomic signature. The signatures from perfect inverted repeats are consistent, therefore, can be used for distinguishing the closely related viruses and differing virus mutation variants. The quantitative comparison of the signature can also provide phylogenetic taxonomy when appropriate numerical metrics for the signatures are realized. Therefore, the perfect inverted repeats can be an effective barcode to delimit species and genotypes.   

\section*{Acknowledgments}
We sincerely appreciate the researchers worldwide who sequenced and shared the complete genome data of SARS-CoV-2 and other coronaviruses from GISAID (https://www.gisaid.org/). This research is partially supported by the National Natural Science Foundation of China (NSFC) grant (91746119, to S.S.-T. Yau), Tsinghua University Spring Breeze Fund (2020Z99CFY044, to S.S.-T. Yau), Tsinghua University start-up fund, and Tsinghua University Education Foundation fund (042202008, to S.S.-T. Yau). 

\section*{Competing interests}
We declare we have no competing interests.

\section*{Abbreviations}
\begin{itemize}	
	\item COVID-19: coronavirus disease 2019 
	\item SARS: severe acute respiratory syndrome
	\item SARS-CoV-2: severe acute respiratory syndrome coronavirus 2
	\item MERS-CoV: Middle East Respiratory Syndrome coronavirus
	\item CRISPR: clusters of regularly interspaced short palindromic repeats
	\item ACE2: angiotensin-converting enzyme 2
	\item NCBI: National Center for Biotechnology Information (USA)
	
\end{itemize}
\clearpage
\bibliographystyle{elsarticle-harv}
\bibliography{../References/myRefs}
\end{document}